\documentclass[11pt]{ieee}
\usepackage{amsmath,concrete,pstricks,epsfig,pst-node}

\usepackage{color,amssymb}
\newcommand{\bm}[1]{\mbox {\boldmath $#1$}}
\newcommand{\bms}[1]{\mbox {\boldmath ${}_{#1}$}}

\newcommand{\ops}[1]{|#1 \rangle\langle #1|}

\newcommand{\bra}[1]{\langle #1|}
\newcommand{\ket}[1]{|#1 \rangle}

\newtheorem{theorem}{Theorem}
\newtheorem{example}{Example}
\newtheorem{lemma}{Lemma}
\newtheorem{definition}{Definition}
\newcommand{\tr}{\operatornamewithlimits{Tr}}
\begin{document}

\thispagestyle{empty}
       \title{\LARGE\rm\bf COMPRESSING MIXED-STATE SOURCES\\[-2mm] BY SENDING CLASSICAL INFORMATION\\[20mm]}
\author{
Emina Soljanin\\[-1.5mm]
Bell Labs, Lucent\\[-1.5mm]
Rm. 2C-177, 600 Mountain Av.\\[-1.5mm]
Murray Hill NJ 07974\\[-1.5mm]
USA\\[-1.5mm]
emina@lucent.com\\[20mm]
}

\maketitle
\thispagestyle{empty}

\begin{abstract}
We consider visible compression for discrete memoryless sources of mixed quantum states
when only classical information can be sent from Alice to Bob.
We assume that Bob knows the source statistics, and that Alice and Bob have identical
random number generators.
We put in an information theoretic framework some recent results on 
visible compression for sources of states with commuting density operators, and remove
the commutativity requirement. We derive a general achievable compression rate, which is
for the noncommutative case still higher than the known lower bound.
We also present several related problems of classical
information theory, and show how they can be used to answer some questions of the
mixed state compression problem.

\vspace{0.1in}
\end{abstract}
\vfill

\today
\newpage

\section{Introduction}
A discrete memoryless source (DMS) of information produces a sequence of independent, 
identically distributed
random variables taking values in a finite set called the {\it source alphabet}.
In quantum systems, source letters are mapped into {\it quantum states} for
quantum transmission or storage.
In the simplest case, quantum states correspond to unit length column vectors in a $d$-dimensional 
Hilbert space ${\cal H}_d$. Such quantum states are called {\it pure}.
When $d=2$, quantum states are called {\it qubits}.
A column vector is denoted by $\ket{\varphi}$, its transpose
by $\bra{\varphi}$. A pure state is mathematically described by its {\it density matrix}
equal to the outer product $\ops{\varphi}$.
In a more complex case, a quantum state can be any of a finite number of
possible pure states $\ket{\varphi_i}$ with probability $p_i$. 
Such quantum states are called {\it mixed}. A mixed state is also described by its density matrix
which is equal to $\sum_ip_i\ops{\varphi_i}$. Note that a density matrix is 
a $d\times d$ Hermitian trace-one positive semidefinite matrix.
A classical analog to a mixed state can be a multi-faced coin which turns up as
any of its faces with the corresponding probability. 

Compression algorithms deal with source sequences rather than individual letters.
There are two possible scenarios for which algorithms can be designed: 
{\it visible} when the encoder Alice knows 
the source sequence and {\it blind} when only the quantum state
corresponding to the sequence is available to her.
The quantum state corresponding to a source sequence of length $n$ has 
a $d^n\times d^n$ density matrix,
equal to the tensor product of density matrices corresponding to the letters in the
sequence. In the blind case, lossless quantum compression algorithms map (encode) these product states into
states over Hilbert spaces of smaller dimension 
with arbitrarily high expected reconstruction (decoding) {\it fidelity} as $n\rightarrow \infty$.
Operations used for encoding and decoding have to be allowed by 
quantum mechanics. 
In the visible case, Alice can as well compress the available classical information, 
which the decoder
Bob can use to prepare a quantum state that (as in the blind case) approximate Alice's 
with arbitrarily high expected fidelity as $n\rightarrow \infty$.

The main question asks what the best compression compatible with the
fidelity goal and encoding/decoding constraints for each scenario is.
The answer to the question was given by Schumacher for discrete memoryless sources of pure 
quantum states \cite{sc}.
Lossless compression of sources of possibly mixed quantum states is not yet fully understood,
and is the subject of current research \cite{hor:oc}--\cite{ki:hms}. 
The optimal compression rate for the blind case scenario was found by Koashi and Imoto in \cite{ki:hms}.
A lower bound to the compression rate was established by Horodecki in \cite{hor:cl} and by
Barnum, Caves, Fuchs, Jozsa, and Schumacher in \cite{fm}.
The optimal compression rates for some special cases were found by Horodecki in \cite{hor:oc} and 
by Barnum, Caves, Fuchs, Jozsa, and Schumacher in \cite{fm}.
More recently, 
an algorithm achieving the lower bound to the compression rate 
for the visible case of states with commuting density operators was presented 
by D\"{u}r, Vidal, and Cirac in
\cite{dvc}, and a possibly related classical information theory problem was discussed 
by Kramer and Savari in \cite{ks}.
Some of these results will be addressed in more detail after the problem we are dealing with
is precisely formulated.

We are concerned with visible compression of discrete memoryless sources
when only classical information can be sent from Alice to Bob.
We assume that Bob knows the source statistics, and that Alice and Bob have identical 
random number generators.
This scenario is the one studied by 
D\"{u}r, Vidal, and Cirac for the case of states with commuting density operators
\cite{dvc}. 
When put in an information theoretic framework, the commutativity requirement 
can be easily removed, and an achievable rate can be found in the same manner.
However, the derived achievable rate is still
higher than the lower bound.

In the second part of the paper, we present several related problems of classical
information theory, and show how they can be used to answer some questions of the
mixed state compression problem.
This paper is written for both information theorists and physicists, although papers
written for two audiences often satisfy neither. 
Here writing for these two groups of scientists merely means that we tried to keep
the paper as self contained as possible, and presented proofs and other material 
in an elementary rather than the most efficient way.

\subsection{Problem Formulation}

Let ${\cal X}$ be a finite set (alphabet), and $\{\rho_a| a\in {\cal X}\}$ a set
of (possibly mixed) quantum states in a $d$-dimensional Hilbert space ${\cal H}_d$.
Let ${\cal P}({\cal X})$ be the set of all probability distributions on ${\cal X}$,
and $P\in {\cal P}({\cal X})$ a particular distribution. 
The set ${\cal E} = \{\rho_a, P(a) | a\in {\cal X}\}$
is usually referred to as an ensemble of mixed states indexed by the elements of ${\cal X}$.
The density matrix of the ensemble ${\cal E}$, which we shall also refer to as the source density
matrix, is given by
\begin{equation}
\rho = \sum_{a\in{\cal X}} P(a)\rho_a.
\label{eq:sdm}
\end{equation}
We shall assume that states $\rho_a$ are mixtures of known (possibly nonorthogonal)
pure states as follows:
Let ${\cal Y}$ be a finite set, and $\{|\psi_b\rangle\langle\psi_b| \bigl\lvert b\in {\cal Y}\}$
be a set of pure quantum states in ${\cal H}_d$ indexed by the elements of ${\cal Y}$.
Let $W$ be an $|{\cal X}|\times |{\cal Y}|$ stochastic matrix with elements 
$W_{ab}=W(b|a)$, $a\in {\cal X}$, $b\in {\cal Y}$, where
$W(\cdot|a)$ is a probability distribution on ${\cal Y}$
for each $a\in {\cal X}$.
We assume that no two states $\rho_a$ are identical in the sense that no two rows of
$W$ are identical.
The density matrices in ${\cal E}$ are given by
\begin{equation}
\rho_a = \sum_{b\in{\cal Y}} W(b|a)|\psi_b\rangle\langle\psi_b|, ~ a\in {\cal X}.
\label{dma}
\end{equation}

A source producing mixed states $\rho_a$, $a\in {\cal X}$, independently according to the 
probability distribution $P$, effectively produces pure states 
$\ops{\psi_b}$, $b\in{\cal Y}$, independently according to the probability distribution $Q$:
\[
Q(b) = \sum_{a\in {\cal X}}P(a)W(b|a).
\]
Thus the source density matrix (\ref{eq:sdm}) can also be expressed in 
terms of $\ops{\psi_b}$ and $Q(b)$, $b\in {\cal Y}$:
\[
\rho = \sum_{b\in {\cal Y}} Q(b)\ops{\psi_b}.
\]

\begin{example} A possible mixed state ensemble is shown in Fig.~\ref{fig:mbenz}. Here $d=2$, 
$|{\cal X}| = 2$, and $|{\cal Y}| = 3$.
\begin{figure}[htb]
\begin{center}
\input{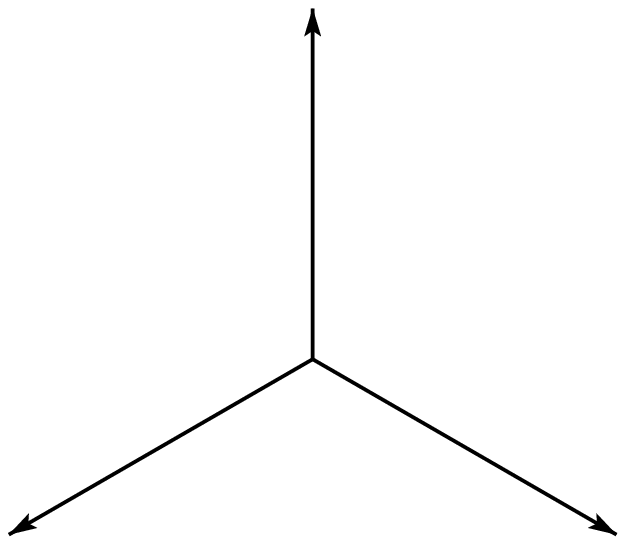}
\end{center}
\caption{A mixed state ensemble.}
\label{fig:mbenz}
\end{figure}
\label{ex:mbenz}
\end{example}

The memoryless source produces sequences of letters,  
where each letter is drawn from the set ${\cal X}$ independently
according to the probability distribution $P$.
Thus a  source sequence
$\bm{x} = (x_1,\dots , x_n) \in {\cal X}^n$ occurs with probability
$P(\bm{x})= P(x_1)\cdot\ldots\cdot P(x_n)$, and the corresponding state has a density
matrix $\bm{\rho}\bms{x} = \rho_{x_1}\otimes\dots \otimes \rho_{x_n}$.
On the transmitting end, the encoder Alice knows ${\cal E}$ and $\bm{x}$.
On the receiving end, the decoder Bob knows ${\cal E}$.
In addition, Alice and Bob have identical random number generators.

For each source sequence $\bm{x}$, Alice prepares and sends to Bob $Rn$ 
bits of classical information,
which he uses (together with his prior knowledge of ${\cal E}$)
to prepare state $\hat{\bm{\rho}}_{\bm{x}}$.
To measure how faithfully mixed state $\sigma$ approximates mixed state $\omega$
and vice versa,
we use the so called {\it mixed state fidelity} $F$ defined as
\begin{equation}
F(\sigma,\omega) = \Bigl\{\tr\bigl[(\sqrt{\sigma}\omega\sqrt{\sigma})^{1/2}\bigr]\Bigr\}^2,
\label{eq:fid}
\end{equation}
whose maximum value is 1.
We shall say that the  mixed state compression is lossless when 
the expected value of
$F(\bm{\rho_x}, \hat{\bm{\rho}}_{\bm{x}})$
can be made arbitrarily close to 1 by increasing the length $n$ of the source sequence:
\begin{equation}
\sum_{\bm{x}\in {\cal X}^n}P(\bm{x})F(\bm{\rho_x}, \hat{\bm{\rho}}{\bms{x}})\rightarrow 1 ~
\text{as} ~ n\rightarrow \infty.
\label{eq:gfc}
\end{equation}

\subsection{Information Measures}
In compression of mixed-state sources by sending classical information,
the well known classical information measures will play a role.
Entropy $H(Q)$, conditional entropy $H(W/P)$, and
mutual information $I(P,W)$ are defined as
\begin{align}
H(Q) = & \; -\sum_{b\in {\cal Y}}Q(b)(a)\log Q(b)\nonumber\\
H(W/P) = & \; -\sum_{a\in {\cal X}}P(a)
\sum_{b\in {\cal Y}}W(b|a)\log W(b|a)
\label{eq:cim}\\
I(P,W) = & \; H(Q) - H(W/P)\nonumber
\end{align}
The corresponding quantum information measures are the source Von Neumann entropy $S(\rho)$, 
the expected value of the Von Neumann entropies
of the source letters $\bar{S}$,
and the Holevo quantity $\chi$:
\begin{align}
S(\rho) = & \; -\tr\rho\log\rho\nonumber\\
\bar{S} = & \;  \sum_{a\in {\cal X}}P(a)S(\rho_a)
\label{eq:qim}\\
\chi = & \; S(\rho) - \bar{S}\nonumber
\end{align}
When $\ops{\psi_b}$, $b\in {\cal Y}$, are orthogonal, the quantum quantities (\ref{eq:qim})
and their classical counterparts (\ref{eq:cim}) are equal:
\begin{align*}
S(\rho) = & \; H(Q)\\
\bar{S} = & \; H(W/P)\\
\chi = & \; I(P,W)
\end{align*}

For the classical information theory problems discussed in Sec.~\ref{sec:cp}, we also need 
stochastic matrix $U$ with elements
$U_{ba}=U(a|b)$, $a\in {\cal X}$, $b\in {\cal Y}$, where
$U(\cdot|b)$ is a probability distribution on ${\cal X}$
for each $b\in {\cal Y}$. The elements of $U$ are computed as
\[
U(a|b) = P(a)W(b|a)/Q(b).
\]
Entropy $H(P)$, conditional entropy $H(U/Q)$, and
mutual information $I(Q,U)$ are defined as the corresponding 
quantities in (\ref{eq:cim}).

\subsection{Known Results}
For sources of pure quantum states, the optimal compression rate
is $S(\rho)$ for both visible and blind scenarios; the information sent from Alice to
Bob is quantum \cite{sc}. For sources of mixed quantum states and 
the fidelity criterion (\ref{eq:gfc}), the following has been shown:
The Von Neumann entropy $S(\rho)$ is the optimal compression rate 
in the blind case scenario \cite{ki:hms}; the compression algorithm is the same
as in the pure case state.
A lower bound to the compression rate of any compression scheme 
is the Holevo quantity $\chi$ \cite{hor:cl}, \cite{fm}.
This lower bound can be achieved by a specific compression algorithm in the
case of quantum states with commuting density operators \cite{dvc};
 the information sent from Alice to Bob is classical.
Achievable compression rates for 
both visible and blind scenarios for sources of quantum states with commuting density operators
and a fidelity criterion different than (\ref{eq:gfc}) are found in \cite{ks}
(see Sec.~\ref{sec:tc}).

When the density matrices $\rho_a$, $a\in {\cal X}$, commute, they can be made diagonal in the
same basis. Thus, one can assume that they are mixtures of orthogonal pure
states $\ops{\psi_b}$, $b\in{\cal Y}$. We address the general case, {\it i.e.,} the
one where the  $\ops{\psi_b}$, $b\in{\cal Y}$, are not necessarily orthogonal.


\newpage
\subsection{The Idea for the Compression Algorithm}
The main idea is simple to state for the reader already familiar with the notion
of typicality as well as the notion of joint and conditional typicality.
A rigorous description, given in the proceeding sections, uses the 
precision provided by the method of types.

For each $\bm{x}$, Alice's state $\bm{\rho_x}$ is roughly a uniform mixture of pure states
$\ops{\Psi\bm{{}_{y}}}=\ops{\psi_{y_1}}\otimes\dots \otimes\ops{\psi_{y_n}}$ where
$\bm{y}$ is conditionally $W$-typical with respect to $\bm{x}$,
and some unlikely pure states.
For each $P$-typical $\bm{x}$, there are about $\exp[nH(W/P)]$ such $\bm{y}$s, and
they are $Q$-typical. There are about $\exp[nH(Q)]$ $Q$-typical $\bm{y}$s, and
a randomly chosen $\bm{y}$ will be conditionally $W$-typical with respect to
any $P$-typical $\bm{x}$ with probability of about $\exp[nH(W/P)]/\exp[nH(Q)]=exp[-nI(P,W)]$.
Therefore, if 
Bob forms a list of $\exp[nI(P,W)]$ randomly chosen 
$Q$-typical $\bm{y}$s, then with high probability there will be a conditionally $W$-typical
$\bm{y}$ with respect to any $P$-typical $\bm{x}$ Alice may have.
If Alice and Bob use identical random number generators to form a list, Alice 
(who knows $\bm{x}$) can identify
such $\bm{y}$ to Bob by sending about $nI(P,W)$ bits of classical information. 
Bob can then prepare the corresponding $\ops{\Psi\bm{{}_{y}}}$, or an error
state if no  $W$-typical $\bm{y}$ was on the list.
Therefore, for every $P$-typical $\bm{x}$, Bob's
state $\hat{\bm{\rho}}\bms{x}$ is with high probability also a uniform mixture 
of pure states $\ops{\Psi\bm{{}_{y}}}$ where
$\bm{y}$ is conditionally $W$-typical with respect to $\bm{x}$
and an unlikely error state.

The idea relies on Shannon's famous observation that ``it is possible for most
purposes to treat long sequences as though there were just $2^{Hn}$ of them,
each with probability $2^{-Hn}$'' \cite{sh:mtc}.
The limitations of this ``typical sequence'' approach becomes apparent when 
one realizes how stringent
requirement the fidelity (\ref{eq:fid}) is. For probability distributions
(diagonal density matrices), the
fidelity is essentially equivalent to the $L_1$ distance (see for example 
\cite[Ch.~9]{cn}). 
In the scheme sketched above, every sequence on Bob's list of randomly chosen
$Q$-typical $\bm{y}$s appears with exactly the same probability.
Bob's
state $\hat{\bm{\rho}}\bms{x}$ is with high probability a uniform mixture
of pure states $\ops{\Psi\bm{{}_{y}}}$, where $\bm{y}$ is conditionally $W$-typical with 
respect to $\bm{x}$.
Alice's state $\bm{\rho}\bms{x}$, is also with high probability a mixture of the same
pure states  $\ops{\Psi\bm{{}_{y}}}$, but not exactly uniform.

Thus for formal proofs, we use a simple refinement of the method
of typical sequences, known as the method of types \cite{Csiszar98}, \cite{CK}.
Two sequences over some alphabet ${\cal A}$  have the same type if 
each letter in ${\cal A}$ appears in both of them the same number of times.
All sequences of the same type form a type class.
We partition the set of typical sequences into type classes.
Sequences of the same type are equiprobable for a DMS, and Bob can form a
list of sequences randomly chosen from the same type class.
Now he will be dealing with a single type class at the time rather than
the entire set of typical sequences. 
He has to know which type class to choose, but Alice can send that information
to him at no cost to the compression rate asymptotically since the number 
of type classes is polynomial in $n$.
An additional benefit of using the method of types will be the
speed of convergence to 1 of the fidelity when $n\rightarrow\infty$.
When two or more sets of sequences are involved
(as ${\cal X}^n$ and  ${\cal Y}^n$ above), joint and conditional types
have to be considered. 
\newpage
\section{Fidelity of Mixed Quantum States}
\subsection{Fidelity and Trace Distance}
Besides computing the mixed state fidelity (\ref{eq:fid}), one can
measure how close state $\sigma$ is to state $\omega$ by computing the
{\it trace distance} 
\[
D(\sigma,\omega) = \frac{1}{2}\tr|\sigma-\omega|.
\]
Here $|A|$ denotes the positive square root of $A^{\dagger} A$, {\it i.e.,} 
$|A|=\sqrt{A^{\dagger} A}$.
The trace distance and the fidelity are closely related and the following holds:
\begin{equation}
1-F(\sigma,\omega)\le D(\sigma,\omega)\le \sqrt{1-F(\sigma,\omega)^2}.
\label{eq:ftd}
\end{equation}
The trace distance is a metric on the space of density operators, and
therefore the triangle inequality is true: 
\begin{equation}
D(\sigma,\omega)\le  D(\sigma,\tau) + D(\tau,\omega).
\label{eq:tie}
\end{equation}
It has some other useful properties, as well.
When we need one of those properties, we shall switch from the fidelity to the trace
distance and back by making use of the inequalities (\ref{eq:ftd}).

Since we shall have to estimate the trace distance of a mixture of inputs,
the following property, known as {\it strong convexity,} will be useful: 
Let $\{p_i\}$ and $\{q_i\}$ be probability distributions over some index set, and $\omega_i$
and $\sigma_i$ density operators also indexed by the same index set. Then
\begin{equation}
D\Bigl(\sum_i p_i\omega_i, \sum_i q_i\sigma_i\Bigr)\le D(\{p_i\},\{q_i\}) 
+\sum_i  p_iD(\omega_i,\sigma_i).
\label{eq:sctd}
\end{equation}
From strong convexity, it directly follows that the trace distance is jointly convex in its arguments:
\begin{equation}
D\Bigl(\sum_i p_i\omega_i, \sum_i p_i\sigma_i\Bigr)\le 
\sum_i p_iD(\omega_i,\sigma_i).
\label{eq:jctd}
\end{equation}
All the above properties of the mixed state fidelity and trace distance and some additional
are discussed in the excellent survey \cite[Ch.~9]{cn}.

\subsection{Approximating Density Matrices}
The objective of the compression algorithm described in Sec.~\ref{sec:ca} is to leave Bob with states
that faithfully approximate Alice's. Only two 
types of approximations will be used, which we can
already demonstrate by just
using the above properties of the fidelity and the trace distance.

Let $\sigma$ and $\sigma_e$ be two density matrices,
$p_{e,n}$ a sequence of numbers such that $p_{e,n}\rightarrow 0$ as $n\rightarrow \infty$,
and $\omega_n$ defined as follows:
\[
\omega_n = p_{e,n}\sigma_e + (1-p_{e,n})\sigma.
\]
\begin{lemma}
Let $\sigma$ and $\omega_n$ be as defined above. Then
$F(\sigma,\omega_n)\rightarrow 1$ as  $n\rightarrow \infty$.
\label{th:fit2}
\end{lemma}

\begin{proof}
By properties (\ref{eq:ftd}), and strong convexity of the trace distance
(\ref{eq:sctd}), we have
\begin{align*}
F(\sigma,\omega_n) & \ge 1-D(\sigma,\omega_n)\\
& \ge 1 - \frac{1}{2}\bigl\lvert 0-p_{e,n}\bigr\rvert - \frac{1}{2}
\bigl\lvert 1-(1-p_{e,n})\bigr\rvert - D(\sigma,\sigma)\\
& \ge 1 - p_{e,n}.
\end{align*}
\end{proof}

Let ${\cal Y}$ be a finite set and $\pi_n\in {\cal P}({\cal Y}^n)$ a probability distribution 
on ${\cal Y}^n$. Let
$\{\sigma\bms{y}, \pi_n(\bm{y}) \bigr\rvert \bm{y}\in {\cal Y}^n\}$ be an ensemble of (possibly
mixed) states over Hilbert space ${\cal H}_d^{\otimes n}$.
Consider the following density matrix $\sigma_n$: 
\[
\sigma_n=
\sum_{\bm{y}\in {\cal Y}^n}\pi_n(\bm{y})\sigma\bms{y}.
\]
Let ${\cal B}_n\subseteq {\cal Y}^n$ be a probabilistically large set:
$\pi_n({\cal B}_n) = 1 -\epsilon_n$, where $\epsilon_n\rightarrow 0$ as  $n\rightarrow \infty$.
It is intuitively clear that if we replace states $\sigma\bms{y}$, $\bm{y}\in {\cal Y}^n\setminus {\cal B}_n$,
in the expression
for $\sigma_n$ by a fixed state $\sigma_e$,
we obtain a density matrix which 
faithfully represents $\sigma_n$ in the sense of (\ref{eq:fid}) 
when $n\rightarrow\infty$. To prove a slightly stronger result
(which we shall use in Sec.~\ref{sec:bm}), we proceed as follows.

Consider
\begin{align*}
\sigma_n = & \sum_{\bm{y}\in {\cal Y}^n}\pi_n(\bm{y})\sigma\bms{y}\\
= & 
\sum_{\bm{y}\in {\cal B}_n} \pi_n(\bm{y})\sigma\bms{y}+
\sum_{\bm{y}\in {\cal Y}^n\setminus {\cal B}_n} \pi_n(\bm{y})\sigma\bms{y}
\end{align*}
Let $p_{e,n}$ be a sequence of numbers such that $p_{e,n}\rightarrow 0$ as $n\rightarrow \infty$,
and $\hat{\sigma}\bms{y}$, $\bm{y}\in {\cal Y}^n$, a set of density matrices such that
$D(\sigma\bms{y}, \hat{\sigma}\bms{y})\le p_{e,n}$ for all $\bm{y}$.
We define a density matrix $\omega_{n}$ as
\[
\omega_{n} = \sum_{\bm{y}\in {\cal B}_n} \pi_n(\bm{y})\hat{\sigma}\bms{y}+
\sum_{\bm{y}\in {\cal Y}^n\setminus {\cal B}_n} \pi_n(\bm{y})\sigma_e
\]
\begin{lemma}
Let $\sigma_n$ and $\omega_{n}$ be as defined above. Then
$F(\sigma_n,\omega_{n})\rightarrow 1$ as  $n\rightarrow \infty$.
\label{th:fit1}
\end{lemma}

\begin{proof}
By properties (\ref{eq:ftd}), and joint convexity of the trace distance
(\ref{eq:jctd}), we have
\begin{align*}
F(\sigma_n,\omega_{n}) & \ge 1-D(\sigma_n,\omega_{n})\\
& \ge 1 - \sum_{\bm{y}\in {\cal B}_n} \pi_n(\bm{y})D(\sigma\bms{y},\hat{\sigma}\bms{y})+
\sum_{\bm{y}\in {\cal Y}^n\setminus{\cal B}_n} \pi_n(\bm{y})D(\sigma\bms{y},\sigma_e)
\\
& \ge 1-(1-\epsilon_n/2)p_{e,n} - \epsilon_n >
1-p_{e,n} - \epsilon_n.
\end{align*}
\end{proof}

\section{The Method of Types}
\subsection{Types and Typical Sequences}

Let, as before, ${\cal X}$ be a finite set
and ${\cal P}({\cal X})$ the set of all probability distributions on ${\cal X}$.
Given a sequence $\bm{x} = \{x_1,\dots , x_n\} \in {\cal X}^n$ and a letter 
$a\in {\cal X}$, let $N(a|\bm{x})$ denote the number occurrences of $a$ in $\bm{x}$.

\begin{definition}
The {\it type} of a sequence $\bm{x}\in {\cal X}^n$ is the distribution $P\bms{x}\in {\cal P}({\cal X})$ 
given by
\[
P\bms{x}(a) = \frac{1}{n}N(a|\bm{x})~~ \text{for every}~~ a\in {\cal X}.
\]
Conversely, the {\it type class} of 
a distribution $P\in {\cal P}({\cal X})$ is the set ${\sf T}_P^n$ of all sequences of 
type $P$ in ${\cal X}^n$:
\[
{\sf T}_P^n = \{\bm{x}: \bm{x}\in {\cal X}^n ~\text{and} ~ P\bms{x} = P\}.
\] 
\end{definition}
\vspace{3mm}
The subset of ${\cal P}({\cal X})$ consisting of the possible types of sequences $\bm{x}\in {\cal X}^n$
is denoted by ${\cal P}_n({\cal X})$. It is easy to show by elementary combinatorics that
\[
|{\cal P}_n({\cal X})| = \binom{n+|{\cal X}|-1}{|{\cal X}|-1}\le (n+1)^{|{\cal X}|}.
\]
Therefore, there is only a polynomial (in $n$) number of types.

The size of ${\sf T}_P^n$ 
can be bounded as follows:
\begin{lemma} \cite[pp.~30]{CK}
For any type $P\bms{x}$ of sequences in ${\cal X}^n$ 
\[
(n+1)^{-|{\cal X}|}\exp\{nH(P\bms{x})\}\le
|{\sf T}_{P\bms{x}}|\le
\exp\{nH(P\bms{x})\}.
\]
\label{le:txs}
\end{lemma}

\begin{definition}
For any distribution $P$ on  ${\cal X}$, a sequence $\bm{x}\in {\cal X}^n$ is $P$-{\it typical}
with constant $\delta$ if
\[
\Bigl|\frac{1}{n}N(a|\bm{x}) - P(a)\Bigr|\le \delta ~~ \text{for every}~~ a\in {\cal X},
\]
and no $a\in {\cal X}$ with $P(a)=0$ occurs in $x$. The set of such sequences will be
denoted by ${\sf T}_{P,\delta}^n$, and the set of their types by ${\cal P}_n^{P,\delta}({\cal X})$.
\end{definition}
\vspace{3mm}

\begin{lemma} \cite[p.~34]{CK}
For any distribution $P$ on  ${\cal X}$, we have
\begin{equation}
P^{n}({\sf T}_{P,\delta}^n)\ge 1-\frac{|{\cal X}|}{4n\delta^2}.
\label{eq:mz}
\end{equation}
\label{le:tm1}
\end{lemma}
\vspace{3mm}

\subsection{Joint and Conditional Types}
If ${\cal X}$ and ${\cal Y}$ are two finite sets, the {\it joint type} of a pair of
sequences $\bm{x}\in {\cal X}^n$ and $\bm{y}\in {\cal Y}^n$ is defined as a type of the sequence
$\{(x_1,y_1),\dots , (x_n,y_n)\}\in{\cal X}\times {\cal Y}$.
Namely, it is the distribution $P\bms{x}_,\bms{y}\in {\cal P}({\cal X}\times {\cal Y})$ given by
\[
P\bms{x}_,\bms{y}(a,b) = \frac{1}{n}N(a,b|\bm{x},\bm{y})~~ \text{for every}~~ a\in {\cal X}, ~ b \in {\cal Y}.
\]
Joint types are often given in terms of the type of $\bm{x}$ and a stochastic matrix
$V:  {\cal X}\rightarrow {\cal Y}$ as
\[
P\bms{x}_,\bms{y}(a,b) = P_{\bm{x}}(a)V(b|a) ~~ \text{for every}~~ a\in {\cal X}, ~ b \in {\cal Y}.
\]

\begin{definition}
We say that $\bm{y}\in {\cal Y}^n$ has 
{\it conditional type} $V$ given $\bm{x}\in {\cal X}^n$ if 
\[
N(a,b|\bm{x},\bm{y}) = N(a|\bm{x})V(b|a) ~~ \text{for every}~~ a\in {\cal X}, ~ b \in {\cal Y}.
\]
For any given $\bm{x}\in {\cal X}^n$ and a stochastic matrix $V: {\cal X}\rightarrow {\cal Y}$, 
the set of sequences $\bm{y}\in {\cal Y}^n$
having conditional type $V$ given $\bm{x}$ is called $V$-{\it shell} of $\bm{x}$, and is denoted
by ${\sf T}_V^n(\bm{x})$ or simply by ${\sf T}_V(\bm{x})$.
The set of all conditional types of $\bm{y}\in {\cal Y}$ for a given $\bm{x}$ 
will be denoted by ${\cal V}_n({\cal Y}, \bm{x})$.
\end{definition}
\vspace{3mm}

The size of a $V$-shell 
can be bounded as follows:
%
\begin{lemma} \cite[pp.~31]{CK}
For any type $P\bms{x}$ of sequences in ${\cal X}^n$ and stochastic matrix $V$ such
that ${\sf T}_V(\bm{x})$ is not empty:
\[
(n+1)^{-|{\cal X}||{\cal Y}|}\exp\{nH(V|P\bms{x})\}\le
|{\sf T}_V(\bm{x})|\le
\exp\{nH(V|P\bms{x})\}.
\]
\label{le:cts}
\end{lemma}

Clearly, every $\bm{y}$ in the $V$-shell of an $\bm{x}$ in the type class ${\sf T}_{P\bms{x}}^n$
has the same type $P\bms{y}$: 
\[
P\bms{y}(b) = \sum_{a\in {\cal X}} P_{\bm{x}}(a)V(b|a).
\]
However, by Lemmas~\ref{le:txs} and \ref{le:cts}, we immediately see that
${\sf T}_V(\bm{x})$ is ``exponentially smaller'' than ${\sf T}_P$, unless all
rows of $V$ are equal to $P\bms{y}$:
\begin{equation}
(n+1)^{-|{\cal X}||{\cal Y}|}\exp\{-nI(P\bms{x},V)\} 
\le \frac{|{\sf T}_V(\bm{x})|}{|{\sf T}_{P\bms{y}}|}\le
(n+1)^{|{\cal Y}|}\exp\{-nI(P\bms{x},V)\}.
\label{eq:ipv}
\end{equation}

\begin{definition}
For any given $\bm{x}\in {\cal X}^n$ and a stochastic matrix $W: {\cal X}\rightarrow {\cal Y}$, 
sequence $\bm{y}\in {\cal Y}^n$ is $W$-generated by $\bm{x}$
(or $W$-typical under the condition $\bm{x}$) with constant $\delta^\prime$ if
\[
\Bigl|\frac{1}{n}N(a,b|\bm{x},\bm{y}) - \frac{1}{n}
 N(a|\bm{x})W(b|a)\Bigr|\le \delta^\prime ~~ \text{for every}~~ 
a\in {\cal X},  ~ b \in {\cal Y},
\]
and $N(a,b|\bm{x},\bm{y})=0$ whenever $W(b|a)=0$. The set of such sequences will be
denoted by ${\sf T}_{W,\delta^\prime}^n(\bm{x})$, and the set of their conditional types by 
${\cal V}_n^{W,\delta^\prime}({\cal Y}, \bm{x})$.
\label{def:xgen}
\end{definition}
\vspace{3mm}
\begin{lemma} \cite[p.~34]{CK}
For any stochastic matrix $W: {\cal X}\rightarrow {\cal Y}$, we have
\[
W^n({\sf T}_{W,\delta^\prime}^n(\bm{x})|\bm{x})=1-\frac{|{\cal X}||{\cal Y}|}{4n\delta^{\prime 2}}.
\]
\label{le:tm2}
\end{lemma}

\subsection{Conditional Typical States}
Let $\rho_a$ be the density matrix of mixed state $a$ given by
(\ref{dma}). 
We consider $\bm{\rho}{\bms{x}} =  \rho_{x_1}\otimes\dots \otimes \rho_{x_n}$ for
$\bm{x}\in {\sf T}_{P\bms{x}}^n$:
\begin{align*}
\bm{\rho}\bm{{}_{x}} = &
\Bigl(\sum_{b\in{\cal Y}} W(b|x_1)\ops{\psi_b}\Bigr)
\otimes\dots \otimes
\Bigl(\sum_{b\in{\cal Y}} W(b|x_n)\ops{\psi_b}\Bigr)\\
= & \sum_{\bm{y}\in{\cal Y}^n} W(y_1|x_1)\cdot\ldots\cdot W(y_n|x_n)\ops{\psi_{y_1}}\otimes\dots
\otimes\ops{\psi_{y_n}}\\
= & \sum_{\bm{y}\in{\cal Y}^n}W^n(\bm{y}|\bm{x})\ops{\Psi\bm{{}_{y}}},
\end{align*}
where $W^n(\bm{y}|\bm{x})$ denotes $W(y_1|x_1)\cdot\ldots\cdot W(y_n|x_n)$ and
$\ops{\Psi\bm{{}_{y}}}$ denotes $\ops{\psi_{y_1}}\otimes\dots \otimes\ops{\psi_{y_n}}$.
We define partial density matrices $\bm{\rho}\bms{x}(V)$ corresponding to each $V$-shell in
${\cal V}_n({\cal Y},\bm{x})$ as follows:
\[
\bm{\rho}\bms{x}(V)=\sum_{\bm{y}\in {\sf T}_V(\bm{x})}\frac{1}{|{\sf T}_V(\bm{x})|}\ops{\Psi\bms{y}}.
\]
Now we can write $\bm{\rho}\bm{{}_{x}}$ as
\begin{equation}
\bm{\rho}\bms{x} = \sum_{V\in {\cal V}_n^{W,\delta^\prime}({\cal Y},\bm{x})}W^n({\sf T}_V(\bm{x})|\bm{x})
\bm{\rho}\bms{x}(V)
+\sum_{V\in {\cal V}_n({\cal Y},\bm{x})\setminus {\cal V}_n^{W,\delta^\prime}({\cal Y},\bm{x})}
W^n({\sf T}_V(\bm{x})|\bm{x})\bm{\rho}\bms{x}(V),
\label{eq:arx}
\end{equation}
where the first term includes only the conditionally typical $V$-shells ($W$-generated
by $\bm{x}$), and 
the second term takes care of the rest.

We are now ready to describe a mixed state compression algorithm. We shall see that
for every typical $\bm{x}$, the algorithm leaves Bob
with mixed state $\hat{\bm{\rho}}\bms{x}$ which differs from Alice's $\bm{\rho}\bm{{}_{x}}$ of (\ref{eq:arx})
only in the following:
In the first term of (\ref{eq:arx}), $\bm{\rho}\bms{x}(V)$ is approximated by $\hat{\bm{\rho}}\bms{x}(V)$ 
in the sense of Lemma~\ref{th:fit2}, in the second term of (\ref{eq:arx}), $\bm{\rho}\bms{x}(V)$
is simply replaced by some fixed error-state $\rho_{e,}\bms{x}$.
Consequently, $\bm{\rho}\bms{x}$ is approximated by $\hat{\bm{\rho}}\bms{x}$ in the sense of Lemma~\ref{th:fit1}.

\section{Mixed State Compression\label{sec:msc}}

\subsection{The Algorithm\label{sec:ca}}

Alice and Bob have identical random number generators.

\begin{enumerate}
\item Alice is given a visible source sequence $\bm{x}\in {\cal X}^n$.
\item For every $a \in {\cal X}$, Alice determines $N(a|\bm{x})$, {\it i.e.} the type
$P\bms{x}$.
\item 
If $P\bms{x}$ is not in ${\cal P}_n^{P,\delta}({\cal X})$, {\it i.e.,}
$\bm{x}$ is not $P$-typical with constant $\delta$, 
Alice sends an error indicator, and Bob prepares some fixed error-state
$\rho_e$. Otherwise, 
they proceed as follows:
\item Alice chooses a conditional type for sequence $\bm{y}$, say $V$,
at random with probability $W^n({\sf T}_V(\bm{x}))$. 
If $V$ is not in ${\cal V}_n^{W,\delta^\prime}({\cal Y}, \bm{x})$, {\it i.e.,}
$\bm{y}$ is not $W$-generated by $\bm{x}$ with constant $\delta^\prime$, 
Alice sends and error indicator, and Bob prepares some fixed error-state
$\rho_{e,}\bms{x}$. 
(Here $\rho_{e,}\bms{x}$ and $\rho_{e,}\bms{x}(V)$ below do not depend on
$\bm{x}$ and $V$ since Bob does not have that information. The notation signifies the stage 
in the algorithm). 
Otherwise, they proceed as follows:
\item Alice determines type $P\bms{y}$ by computing
\[
P\bms{y}(b) = \sum_{a\in {\cal X}} P_{\bm{x}}(a)V(b|a).
\]
\item Alice tells the type $P\bms{y}$ to Bob by sending $\log |{\cal P}_n({\cal Y})|$ bits
identifying the particular $P\bms{y}$. 
\item Alice and Bob each form a list of $N_l$ sequences $\bm{y}$ by drawing
randomly from the type class ${\sf T}_{P\bms{y}}$. Let
\[
R=\frac{\log N_l}{n}.
\]
\item If there is one or more $\bm{y}$'s on the list belonging to the $V$-shell ${\sf T}_V(\bm{x})$,
Alice sends $\log N_l$ bits to Bob identifying the position of first $\bm{y}\in {\sf T}_V(\bm{x})$
on the list, and Bob prepares $\ops{\Psi\bm{{}_{y}}}$.  
With some probability $p_{e,}\bms{x}(V)$, no
$\bm{y}\in {\sf T}_V(\bm{x})$ will be on the list that Alice and Bob form.
If that is the case, Alice sends an error indicator and Bob prepares some fixed error-state
$\rho_{e,}\bms{x}(V)$.
\end{enumerate}

\subsection{Bob's Density Matrix}
For non-typical $\bm{x}$, Bob's state is $\rho_e$, 
while for typical $\bm{x}$, his state is given by
\begin{align*}
\hat{\bm{\rho}}\bm{{}_{x}} = & 
\sum_{V\in {\cal V}_n^{W,\delta^\prime}({\cal Y},\bm{x})}W^n(({\sf T}_V(\bm{x})|\bm{x})\hat{\bm{\rho}}\bms{x}(V)+\\
& \sum_{V\in {\cal V}_n({\cal Y},\bm{x})\setminus {\cal V}_n^{W,\delta^\prime}({\cal Y},\bm{x})}W^n(({\sf T}_V(\bm{x})|\bm{x})
\rho_{e,}\bms{x},
\qquad \bm{x}\in{\sf T}^n_{P,\delta}.
\end{align*}
Here $\hat{\bm{\rho}}\bms{x}(V)$ denotes Bob's density matrix when conditional type $V$ is chosen by Alice.
Since Bob prepares either the error-state $\rho_{e,}\bms{x}(V)$ with probability $p_{e,}\bms{x}(V)$, or one of
the states $\ops{\Psi\bm{{}_{y}}}$, $\bm{y}\in {\sf T}_V(\bm{x})$, with probability $1-p_{e,}\bms{x}(V)$,
we have
\begin{align*}
\hat{\bm{\rho}}\bms{x}(V)= & 
p_{e,}\bms{x}(V)\rho_{e,}\bms{x}(V)+
(1-p_{e,}\bms{x}(V))\sum_{\bm{y}\in {\sf T}_V(\bm{x})}\frac{1}{|{\sf T}_V(\bm{x})|}\ops{\Psi\bms{y}}.
\end{align*}
Note that if $p_{e,}\bms{x}(V)\rightarrow 0$ as $n\rightarrow \infty$, then Bob's $\hat{\bm{\rho}}\bms{x}(V)$
approximates Alice's $\bm{\rho}\bms{x}(V)$ in the sense of Lemma~\ref{th:fit2}, and 
thus Bob's $\hat{\bm{\rho}}\bm{{}_{x}}$
approximates Alice's $\bm{\rho}\bm{{}_{x}}$ in the sense of Lemma~\ref{th:fit1}.

To see under which conditions $p_{e,}\bms{x}(V)\rightarrow 0$ as $n\rightarrow \infty$,
we proceed as follows:
Clearly, the probability that a sequence $\bm{y}$ randomly drawn from ${\sf T}_{P\bms{y}}$ is in 
${\sf T}_V(\bm{x})$ equals to $|{\sf T}_V(\bm{x})|/|{\sf T}_{P\bms{y}}|$. 
The probability $p_{e,}\bms{x}(V)$ that
no such sequence is on the list of length $N_l$ is thus equal to 
$\bigl(1-|{\sf T}_V(\bm{x})|/|{\sf T}_{P\bms{y}}|\bigr)^{N_l}$.
This quantity can be bound by applying the inequality $(1-x)^k \le e^{-kx}$,
and then the ratio $|{\sf T}_V(\bm{x})|/|{\sf T}_{P\bms{y}}|$ can be bound by
applying the inequalities (\ref{eq:ipv}):
\begin{align*}
p_{e,}\bms{x}(V) = & \bigl(1-|{\sf T}_V(\bm{x})|/|{\sf T}_{P\bms{y}}|\bigr)^{N_l}\\
\le & e^{-N_l|{\sf T}_V(\bm{x})|/|{\sf T}_{P\bms{y}}|}\\
\le & e^{-\exp(n(R-I-\epsilon_n^{\prime\prime}))},
\end{align*}
where $I$ refers to $I(P\bms{x},V)$ and
$\epsilon_n^{\prime\prime}= |{\cal X}||{\cal Y}|\log(n+1)/n$. 
Therefore, if 
\begin{equation}
R > I(P\bms{x},V)+\epsilon_n^{\prime\prime},
\label{eq:rate}
\end{equation}
we have $p_{e,}\bms{x}(V)\rightarrow 0$ as $n\rightarrow \infty$.

\subsection{Mixed State Fidelity\label{sec:bm}}
We now have all we need to bound the value of 
$\sum_{\bm{x}\in {\cal X}^n}P(\bm{x})F(\bm{\rho_x}, \hat{\bm{\rho}}{\bms{x}})$,
and thus prove the main result of the compression algorithm:
\begin{theorem}
Let $R > I(P\bms{x},V)+\epsilon_n^{\prime\prime}$, for all $P\bms{x}\in {\cal P}_n({\cal Y})$
and all $V\in {\cal V}_n^{W,\delta^\prime}({\cal Y}, \bm{x})$. Then
\[
\sum_{\bm{x}\in {\cal X}^n}P(\bm{x})F(\bm{\rho_x}, \hat{\bm{\rho}}{\bms{x}})\rightarrow 1 ~
\text{as} ~ n\rightarrow \infty.
\]
\label{th:cr}
\end{theorem}
\begin{proof}
By Lemma~\ref{le:tm1},
\begin{align}
\sum_{\bm{x}\in {\cal X}^n}P(\bm{x})F(\bm{\rho_x}, \hat{\bm{\rho}}{\bms{x}}) &
\ge 1 - \sum_{\bm{x}\in {\cal X}^n}P(\bm{x})D(\bm{\rho_x}, \hat{\bm{\rho}}{\bms{x}})\nonumber\\
& = 1 - \sum_{\bm{x}\in {\sf T}^n_{P,\delta}}P(\bm{x})D(\bm{\rho_x}, \hat{\bm{\rho}}{\bms{x}})
-\sum_{\bm{x}\in {\cal X}^n\setminus {\sf T}^n_{P,\delta}}P(\bm{x})D(\bm{\rho_x}, \rho_e)\nonumber\\ 
& \ge 1- \sum_{\bm{x}\in {\sf T}^n_{P,\delta}}P(\bm{x})D(\bm{\rho_x}, \hat{\bm{\rho}}{\bms{x}})
-\epsilon_n,
\label{eq:fs1}
\end{align}
where $\epsilon_n = |{\cal X}|/(2n\delta^2)$.
By Lemma~\ref{le:tm2},
\begin{align}
D(\bm{\rho}\bms{x},\hat{\bm{\rho}}\bms{x}) & =
\sum_{V\in {\cal V}_n^{W,\delta^\prime}({\cal Y},\bm{x})}W^n(({\sf T}_V(\bm{x})|\bm{x})
D(\bm{\rho}\bms{x}(V),\hat{\bm{\rho}}\bms{x}(V))
+\sum_{V\in {\cal V}_n({\cal Y},\bm{x})\setminus {\cal V}_n^{W,\delta^\prime}({\cal Y},\bm{x})}W^n(({\sf T}_V(\bm{x})|\bm{x})
D(\bm{\rho}\bms{x}(V),\hat{\bm{\rho}}_{e,}\bms{x})\nonumber\\
& \le \sum_{V\in {\cal V}_n^{W,\delta^\prime}({\cal Y},\bm{x})}W^n(({\sf T}_V(\bm{x})|\bm{x})
D(\bm{\rho}\bms{x}(V),\hat{\bm{\rho}}\bms{x}(V))
+\epsilon_n^\prime,
\label{eq:fs2}
\end{align}
where $\epsilon_n^\prime = |{\cal X}||{\cal Y}|/(2n\delta^{\prime2})$.
By Lemma~\ref{th:fit2},
\begin{equation}
D(\bm{\rho}\bms{x}(V),\hat{\bm{\rho}}\bms{x}(V))\le p_{e,}\bms{x}(V).
\label{eq:fs3}
\end{equation}

Let $p_{e,n}$ denote the maximum of all $p_{e,}\bms{x}(V)$ over all $\bm{x}\in {\sf T}^n_{P,\delta}$
and $V\in {\cal V}_n^{W,\delta^\prime}({\cal Y},\bm{x})$.
Combining (\ref{eq:fs1}), (\ref{eq:fs2}), and (\ref{eq:fs3}),  we obtain
\begin{align*}
\sum_{\bm{x}\in {\cal X}^n}P(\bm{x})F(\bm{\rho_x}, \hat{\bm{\rho}}{\bms{x}})
&\ge 1- (1-\epsilon_n)(1-\epsilon_n^\prime)p_{e,n}-(1-\epsilon_n)\epsilon_n^\prime -\epsilon_n\\
&> 1-p_{e,n}-\epsilon_n^\prime -\epsilon_n
\end{align*}
As $n\rightarrow \infty$, we know that $\epsilon_n\rightarrow 0$ and $\epsilon_n^{\prime}\rightarrow 0$, 
whereas $p_{e,n}\rightarrow 0$ when the compression rate satisfies (\ref{eq:rate})
for all $\bm{x}\in {\sf T}^n_{P,\delta}$ and $V\in {\cal V}_n^{W,\delta^\prime}({\cal Y},\bm{x})$.
Therefore, under the conditions of the Theorem, we have
\[
\sum_{\bm{x}\in {\cal X}^n}P(\bm{x})F(\bm{\rho_x}, \hat{\bm{\rho}}{\bms{x}})\rightarrow 1 ~
\text{as} ~ n\rightarrow \infty.
\]
\end{proof}

\subsection{Achievable Compression Rate}
To show that a compression rate of $I(P,W)$ is achievable, we use the continuity 
of entropy:
\begin{lemma}
If $\{p_i\}_{i=1}^N$ and $\{q_i\}_{i=1}^N$ are two probability distributions such that
\[
\sum_{i=1}^N |p_i-q_i|\le \theta\le\frac{1}{2},
\]
then 
\[
|H(p_1,\dots,p_N)-H(q_1,\dots,q_N)|\le -\theta\log\frac{\theta}{N}.
\]
\label{le:ce}
\end{lemma}

We show that for all $\bm{x}\in {\sf T}^n_{P,\delta}$ and $V\in {\cal V}_n^{W,\delta^\prime}({\cal Y},\bm{x})$,
\[
|I(P,W)-I(P\bms{x},V)|\rightarrow 0~
\text{as}~ \delta,\delta^\prime\rightarrow 0.
\]
Consider
\begin{equation}
\begin{array}{rcl}
|I(P,W)-I(P\bms{x},V)|  & \le & |H(Q)-H(P\bms{y})| + |H(W|P)-H(V|P\bms{x})|\\
& \le  & |H(Q)-H(P\bms{y})| + |H(W|P)-H(W|P\bms{x})|+|H(W|P\bms{x})-H(V|P\bms{x})|\\
& \le  & - |{\cal X}||{\cal Y}|(\delta+\delta^\prime)\log\bigl[(\delta+\delta^\prime)|{\cal X}|\bigr] + 
\delta\log|{\cal Y}| - |{\cal X}||{\cal Y}|\delta^\prime\log\delta^\prime
\end{array}
\label{eq:eb}
\end{equation}
To bound the first and the third term in (\ref{eq:eb}), we used the continuity of
entropy (Lemma~\ref{le:ce}), and to bound the second term, we used the $\log|{\cal Y}|$ bound on the 
entropy of any distribution over ${\cal Y}$.

\section{Applications}

\subsection{The Example of Fig.~\ref{fig:mbenz}}
We consider the system of Example~\ref{ex:mbenz} as shown in Fig.~\ref{fig:mbenz}.
Let $h$ denote the binary entropy function: 
$h(x)=-x\log(x)-(1-x)\log(1-x)$.

For the classical information measures, we have
\begin{align*}
H(Q) = & \log 3\\
H(P/W) = & h(1/3) = -2/3 + \log 3\\
I(P,W) = & 2/3
\end{align*}

For the quantum information measures, we have
\begin{align*}
S(\rho) = & 1\\
\bar{S} = & h(1/2-\sqrt{3}/6)\\
\chi = & 1-h(1/2-\sqrt{3}/6) = .255\dots
\end{align*}
Note the gap between $I(P,W)$ and $\chi$.

\subsection{Sources of Mixed States with Commuting Density Operators}
When the density matrices $\rho_a$, $a\in {\cal X}$, commute, they can be made diagonal in the
same basis. Thus, we shall assume that they are mixtures of orthogonal pure 
states $\ops{\psi_b}$, $b\in{\cal Y}$:
\[
\rho_a = \sum_{b\in{\cal Y}} W(b|a)|\psi_b\rangle\langle\psi_b|, ~ a\in {\cal X}.
\]
Recall that
\[
\rho = \sum_{a\in{\cal X}} P(a)\rho_a = \sum_{b\in{\cal Y}} Q(b)\ops{\psi_b}.
\]
Since $\ops{\psi_b}$ are orthogonal, we have $S(\rho) = H(Q)$, and 
$\sum_{a\in {\cal X}}P(a)S(\rho_a)=H(W/P)$. Therefore, the Holevo quantity
$\chi$ is in this case equal to the mutual information $I(P,W)$:
\[
\chi = S(\rho) - \sum_{a\in {\cal X}}P(a)S(\rho_a)=H(Q)-H(W/P) = I(P,W).
\]

A way to ensure that Bob's matrices $\hat{\bm{\rho}}\bms{x}$ commute
is to assign the uniform mixture of
pure states $\ops{\Psi\bms{y}}$, $\bm{y}\in {\cal Y}^n$, to each error-state 
in the compression algorithm:
\begin{equation}
\rho_e = \rho_{e,}\bms{x}=\rho_{e,n}=
\frac{1}{|{\cal Y}|^n}\sum_{\bm{y}\in {\cal Y}^n} \ops{\Psi\bms{y}}.
\label{eq:esa}
\end{equation}
Of course, no particular choice of the error-states is required if the only goal
is an asymptotically good fidelity.
However, commutativity of Bob's matrices keeps the entire system classical,
makes it easier to derive an expression for the fidelity, and consequently puts us in
a good position to recognize possible related problems of classical information 
theory.

For each sequence $\bm{x}$, Alice's density matrix is
\[
\bm{\rho}\bms{x} = \sum_{\bm{y}\in{\cal Y}^n}W^n(\bm{y}|\bm{x})\ops{\Psi\bm{{}_{y}}}.
\]
With assignment (\ref{eq:esa}), the compression algorithm leaves Bob with 
the density matrix 
\[
\hat{\bm{\rho}}\bms{x} = \sum_{\bm{y}\in{\cal Y}^n}\widehat{W}^n(\bm{y}|\bm{x})\ops{\Psi\bm{{}_{y}}}.
\]
Therefore, the mixed state fidelity between $\bm{\rho}\bms{x}$ and $\hat{\bm{\rho}}\bms{x}$ is
\begin{align*}
F(\bm{\rho}\bms{x}, \hat{\bm{\rho}}\bms{x}) = &
\Bigl\{\tr\bigl[(\sqrt{\bm{\rho}\bms{x}}\hat{\bm{\rho}}\bms{x}\sqrt{\bm{\rho}\bms{x}})^{1/2}\bigr]\Bigr\}^2=
\Bigl\{\tr\bigl[(\bm{\rho}\bms{x}\hat{\bm{\rho}}\bms{x})^{1/2}\bigr]\Bigr\}^2\\
= & \Bigl\{\tr\bigl[\bigr(\sum_{\bm{y}\in {\cal Y}^n}W^n(\bm{y}|\bm{x})\widehat{W}^n(\bm{y}|\bm{x})
\ops{\Psi\bms{y}}\bigl)^{1/2}\bigr]\Bigr\}^2\\
= & \bigr[\sum_{\bm{y}\in {\cal Y}^n} \sqrt{W^n(\bm{y}|\bm{x})\cdot\widehat{W}^n(\bm{y}|\bm{x})}\Bigl]^2.
\end{align*}

\section{Connections with Classical Problems\label{sec:cp}}
We discuss three problems of classical information
theory, each to a certain degree related to the problem of visible mixed state compression.
\subsection{Sources of Probability Distributions\label{sec:ce}}
We consider a discrete memoryless source whose alphabet
is a set of $|{\cal X}|$ coins with $|{\cal Y}|$ faces.
When coin $C_a$ is tossed, face $b$ appears with probability $W(b|a)$, $a\in {\cal X}$,
$b\in {\cal Y}$. 
The source, Alice, produces sequences of coins, {\it i.e.,} probability distributions,
where each coin is drawn independently
according to the probability distribution $P$. A source whose alphabet consists
of two probability distribution is described in the following example:
\begin{example}
A source of two biased coins is shown in Fig.~\ref{fig:bsc}. 
\begin{figure}[htb]
\begin{center}
\input{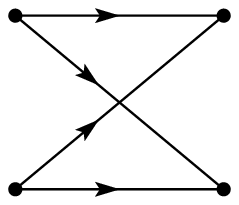}
\end{center}
\caption{Source.}
\label{fig:bsc}
\end{figure}
If coin $C_1$ is tossed, the probability of getting a tail is $w$, if coin $C_2$ is tossed, 
the probability of getting a head is $w$. 
\label{ex:bsc}
\end{example}

When the $n$ coins in Alice's sequence $C\bms{x}=\{C_{x_1},\dots,C_{x_n}\}$ 
are tossed, the probability of getting sequence $\bm{y}\in{\cal Y}^n$ of faces 
is $W^n(\bm{y}|\bm{x})=W(y_1|x_1)\cdot\ldots\cdot W(y_n|x_n)$.
Each time Alice is given sequence of coins $C\bms{x}$, the reproducing source
Bob prepares sequence of faces $\bm{y}\in {\cal Y}^n$ with probability $\widehat{W}^n(\bm{y}|\bm{x})$
such that
\begin{equation}
\sum_{\bm{x}\in {\cal X}^n}P(\bm{x})F_{{\cal Y}^n}
\bigl(W^n(\cdot|\bm{x}),\widehat{W}^n(\cdot|\bm{x})\bigr)\rightarrow 1 ~
\text{as} ~ n\rightarrow \infty.
\label{eq:ca}
\end{equation}
Here $F_{{\cal Y}^n}(\cdot,\cdot)$ is the Bhattacharyya-Wooters overlap between two probability
distributions over the set ${\cal Y}^n$:
\[
F_{{\cal Y}^n}\bigl(W^n(\cdot|\bm{x}),\widehat{W}^n(\cdot|\bm{x})\bigr) = \bigr[
\sum_{\bm{y}\in {\cal Y}^n} \sqrt{W^n(\bm{y}|\bm{x})\cdot\widehat{W}^n(\bm{y}|\bm{x})}\Bigl]^2.
\]

Requirement (\ref{eq:ca}), ensures that
Alice and Bob appear to be identical sources of probability
distributions to an observer who can see only the sequences of faces at both
ends. More precisely, with probability approaching 1 as $n$ increases,
such observer can not tell the difference between Alice and Bob. 
We immediately see that goal (\ref{eq:ca}) can be achieved by running the compression algorithm
described in Sec.~\ref{sec:msc}. 

\subsection{Type Covering\label{sec:tc}}
We again consider the source of the previous section, whose alphabet
is a set of $|{\cal X}|$ coins with $|{\cal Y}|$ faces.
But now, for each Alice's sequence $C\bms{x}$ of coins,
Bob prepares a predetermined sequence $\bm{y}(\bm{x})$ of faces such that
\[
\sum_{\bm{x}\in {\cal X}^n}P(\bm{x})F_{{\cal X}\times {\cal Y}}
\bigl(P\bms{x}W(\cdot|\cdot),
P\bms{x}{}_,\bms{y}{}_(\bms{x}{}_)\bigr) \rightarrow 1 ~
\text{as} ~ n\rightarrow \infty.
\]
Here $F_{{\cal X}\times{\cal Y}}(\cdot,\cdot)$ is the Bhattacharyya-Wooters overlap 
between two probability
distributions over the set ${\cal X}\times{\cal Y}$:
\begin{align}
\label{eq:bw2}
F_{{\cal X}\times {\cal Y}}\bigl(P\bms{x}W(\cdot|\cdot),
P\bms{x}{}_,\bms{y}{}_(\bms{x}{}_)\bigr) =  &
\Bigl[\sum_{(a,b)\in {\cal X}\times{\cal Y}}
\sqrt{P\bms{x}(a)W(b|a)\cdot P\bms{x}{}_,\bms{y}{}_(\bms{x}{}_)(a,b)}\Bigr]^2\\
= & \Bigl[\sum_{(a,b)\in {\cal X}\times{\cal Y}}\frac{1}{n}
\sqrt{N(a|\bm{x})W(b|a)\cdot N(a,b|\bm{x},\bm{y}(\bm{x}))}\Bigr]^2.\nonumber
\end{align}
This problem was translated into a rate distortion one, and solved for perfect
and imperfect asymptotic fidelity in \cite{ks}. 
By using only simple combinatorial techniques, we will show that $I(P,W)$ is the optimal
rate for perfect asymptotic fidelity.

We first show that the overlap (\ref{eq:bw2})
is close to 1
if and only if $\bm{y}(\bm{x})$ is $W$-generated by $\bm{x}$ with 
some constant $\delta$ close to 0. We prove this claim in the following two lemmas
by using the inequalities (\ref{eq:ftd}),
which bound the fidelity in terms of the trace distance and vice versa.


\begin{lemma}
Let $\bm{y}(\bm{x})$ be  $W$-generated by $\bm{x}$ with constant
$\delta$:
\begin{equation}
\Bigl|\frac{1}{n}N(a,b|\bm{x},\bm{y}) - \frac{1}{n}
 N(a|\bm{x})W(b|a)\Bigr|\le \delta ~~ \text{for every}~~
a\in {\cal X},  ~ b \in {\cal Y}.
\label{eq:xgen}
\end{equation}
Then
\vspace{-4mm}
\[
F_{{\cal X}\times {\cal Y}}\bigl(P\bms{x}W(\cdot|\cdot),P\bms{x}{}_,\bms{y}{}_(\bms{x}{}_)\bigr)
\rightarrow 1,~ \text{as}~
\delta \rightarrow 0.
\]
\vspace{0.005mm}
\label{le:xgen}
\end{lemma}

\begin{proof}
By (\ref{eq:ftd}), we can bound the Bhattacharyya-Wooters overlap (\ref{eq:bw2})
in terms of the corresponding trace distance:
\begin{align*}
F_{{\cal X}\times {\cal Y}}\bigl(P\bms{x}W(\cdot|\cdot),
P\bms{x}{}_,\bms{y}{}_(\bms{x}{}_)\bigr) 
= & \Bigl[
\sum_{(a,b)\in {\cal X}\times{\cal Y}}\frac{1}{n}
\sqrt{N(a|\bm{x})W(b|a)\cdot N(a,b|\bm{x},\bm{y}(\bm{x}))}\Bigr]^2\\
\ge &
1 - \frac{1}{2}\sum_{(a,b)\in {\cal X}\times{\cal Y}}\Bigl|\frac{1}{n}N(a,b|\bm{x},\bm{y}) - \frac{1}{n}
N(a|\bm{x})W(b|a)\Bigr|.
\end{align*}
Because of (\ref{eq:xgen}), we have
\[
F_{{\cal X}\times {\cal Y}}\bigl(P\bms{x}W(\cdot|\cdot),
P\bms{x}{}_,\bms{y}{}_(\bms{x}{}_)\bigr) 
\ge
1-\delta\cdot|{\cal X}||{\cal Y}|/2.
\]
\end{proof}

\begin{lemma}
Let Bhattacharyya-Wooters overlap (\ref{eq:bw2}) be equal to $1-\alpha/2$. 
Then sequence $\bm{y}(\bm{x})$ is $W$-generated by $\bm{x}$ with constant $2\sqrt{\alpha}$.
\label{le:xgenc}
\end{lemma}
\begin{proof}
By (\ref{eq:ftd}), we can bound the trace distance between the distributions $P\bms{x}W(\cdot|\cdot)$
and $P\bms{x}{}_,\bms{y}{}_(\bms{x}{}_)$
in terms of their Bhattacharyya-Wooters overlap (\ref{eq:bw2}): 
\[
\frac{1}{2}\sum_{(a,b)\in {\cal X}\times{\cal Y}}\Bigl|\frac{1}{n}N(a,b|\bm{x},\bm{y}) - \frac{1}{n}
N(a|\bm{x})W(b|a)\Bigr| \le 
[1-F^2_{{\cal X}\times {\cal Y}}\bigl(P\bms{x}W(\cdot|\cdot)
P\bms{x}{}_,\bms{y}{}_(\bms{x}{}_)\bigr)]^{1/2}
\le \sqrt{\alpha}.
\]
It follows that
\[
\Bigl|\frac{1}{n}N(a,b|\bm{x},\bm{y}) - \frac{1}{n}
 N(a|\bm{x})W(b|a)\Bigr|\le 2\sqrt{\alpha} ~~ \text{for every}~~
a\in {\cal X},  ~ b \in {\cal Y}.
\]
\end{proof}
Therefore, for a given $\bm{x}$, the fidelity (\ref{eq:bw2})
is close to 1
if and only if $\bm{y}(\bm{x})$ is $W$-generated by $\bm{x}$ with
some constant $\delta$ close to 0. A compression code ${\cal C}\subseteq {\cal Y}^n$ will have to
contain at least one such  $\bm{y}(\bm{x})$ for each 
$\bm{x}\in {\sf T}_{P,\delta\bms{x}}$, as shown next.

\begin{definition}
We shall say that code ${\cal C}\subseteq {\cal Y}^n$ of face sequences 
{\it covers} set ${\cal B} \subseteq {\cal X}^n$ of coin sequences with constant 
$\delta$
if it contains at least one element of ${\sf T}_{W,\delta}^n(\bm{x})$, for each $\bm{x}\in {\cal B}$,
{\it i.e.,} for each $\bm{x}\in {\cal B}$, we have
${\cal C}\cap {\sf T}_{W,\delta}^n(\bm{x})\ne \emptyset$.
\end{definition}

\begin{theorem}
Let ${\cal C}$ be a code which covers the set ${\sf T}_{P,\delta\bms{x}}$.
For each $\bm{x}\in {\cal X}^n$, 
let $\bm{y}(\bm{x})$ be an element of ${\cal C}\cap {\sf T}_{W,\delta}^n(\bm{x})$
if $\bm{x}\in {\sf T}_{P,\delta\bms{x}}$,
and $\bm{y}(\bm{x})$ be an arbitrary $\bm{y}_e\in{\cal Y}^n$, otherwise. Then
\[
\sum_{\bm{x}\in {\cal X}^n}P(\bm{x})F_{{\cal X}\times {\cal Y}}
\bigl(P\bms{x}W(\cdot|\cdot),
P\bms{x}{}_,\bms{y}{}_(\bms{x}{}_)\bigr) \rightarrow 1 ~
\text{as} ~ n\rightarrow \infty.
\]
\label{th:tcpc}
\end{theorem}
\vspace{3mm}

\begin{proof} By Lemmas~\ref{le:tm1} and ~\ref{le:xgen},
\[
\sum_{\bm{x}\in {\cal X}^n}P(\bm{x})F_{{\cal X}\times {\cal Y}}
\bigl(P\bms{x}W(\cdot|\cdot),
P\bms{x}{}_,\bms{y}{}_(\bms{x}{}_)\bigr) \ge
\Bigl(1-\frac{|{\cal X}|}{4n\delta\bms{x}^2}\Bigr)
\Bigl(1-\frac{\delta|{\cal X}||{\cal Y}|}{2}\Bigr).
\]
\end{proof}

We assume that Alice and Bob both know the compression code ${\cal C}$. 
To identify $\bm{y}(\bm{x})$,
Alice has to send to Bob $\log |{\cal C}|$ bits of classical information.
The compression rate is, therefore, given by 
\[
R = \frac{\log |{\cal C}|}{n},
\]
and is determined by the size of the smallest
code ${\cal C}$ that covers ${\sf T}_{P,\delta}$.
To bound the size of ${\cal C}$, we
shall use the following simple general result about coverings, known as
Johnson-Stein-Lov\'{a}sz Theorem
(see for example \cite[p.~322]{chll}):
\begin{theorem}
Let $\bm{A}$ be a $0-1$ matrix with $N$ rows and $M$ columns. Assume that each row
contains at least $v$ ones and each column at most $a$ ones. Then there exists an
$N\times K$ submatrix $\bm{C}$ of $\bm{A}$ with
\[
K\le \frac{N}{a} +\frac{M}{v}\log a \le \frac{M}{v}(1+\log a)
\]
such that $\bm{C}$ contains no all-zero rows.
\label{th:jsl}
\end{theorem}

In order to use Theorem~\ref{th:jsl} in bounding compression code rate $R$,
we construct matrix $\bm{A}$ as follows:
The rows of $\bm{A}$ are indexed by sequences $\bm{x}$ that are $P$-typical with constant
$\delta\bms{x}$, columns by
sequences $\bm{y}$ that are $Q$-typical with constant $\delta\bms{y}$.
Thus $\bm{A}$ has $|{\sf T}_{P,\delta\bms{x}}|$ rows and $|{\sf T}_{Q,\delta\bms{y}}|$ columns.
An element of $\bm{A}$ in row $\bm{x}$ and column $\bm{y}$ is set to $1$ if
$\bm{x}$ and $\bm{y}$ are jointly typical with constant $\delta\bms{xy}$, {\it i.e.,}
if
\[
|P(a)W(b|a)-\frac{1}{n}N(a,b|\bm{x},\bm{y})| = |Q(a)U(b|a)-\frac{1}{n}N(a,b|\bm{x},\bm{y})| 
\le \delta\bms{xy},
\]
otherwise to $0$.
We first show that all $\bm{y}$s corresponding to the 1s in a particular 
row $\bm{x}$ are $W$-generated by $\bm{x}$ with constant $\delta =
\delta\bms{x}+\delta\bms{xy}$:
For each row $\bm{x}$ having a $1$ in column $\bm{y}$, we have
\begin{align*}
|P\bms{x}(b)W(b|a)-\frac{1}{n}N(a,b|\bm{x},\bm{y})| \le &\; |P\bms{x}(b)W(b|a)-P(a)W(b|a)|
+ |P(a)W(b|a)-\frac{1}{n}N(a,b|\bm{x},\bm{y})|\\ 
\le &\;\delta\bms{x}+\delta\bms{xy}= \delta.
\end{align*}
Therefore $\bm{y}\in {\sf T}_{W,\delta}^n(\bm{x})$.
Since $\bm{C}$ is a submatrix of $\bm{A}$
with no all-zero rows, the set ${\cal C}$ of sequences $\bm{y}$ indexing the columns of
$\bm{C}$ covers the set ${\sf T}_{P,\delta\bms{x}}^n$.
Therefore, by Theorem~\ref{th:tcpc}, ${\cal C}$ can serve as a compression code asymptotically 
achieving perfect fidelity.

We find $v$, a lower bound to the number of $1$'s in each row 
as follows:
For each $\bm{x}$, consider all sequences $\bm{y}$ which are $W$-generated by
$\bm{x}$ with constant $\delta^\prime$. If $\delta^\prime$ is set to be equal to
$\delta\bms{xy}-\delta\bms{x}$, we have 
\begin{align*}
|P(a)W(b|a)-\frac{1}{n}N(a,b|\bm{x},\bm{y})| \le & \;
|P(a)W(b|a)-P\bms{x}(a)W(b|a)| + |P\bms{x}(a)W(b|a)-\frac{1}{n}N(a,b|\bm{x},\bm{y})|\\
\le &\;\delta\bms{x} + \delta^\prime = \delta\bms{xy}.
\end{align*}
Thus, if $\bm{y}\in {\sf T}_{W,\delta^\prime}^n(\bm{x})$, the
element of $\bm{A}$ in row $\bm{x}$ and column $\bm{y}$ is set to $1$.
Therefore, the number of $1$'s in each row is at least $v$:
\[
v = \exp[n(H(W/P)-\epsilon^\prime).
\]

We find $a$, an upper bound to the number of $1$'s in each column as follows:
For each column $\bm{y}$ having a $1$ in row $\bm{x}$, we have
\begin{align*}
|P\bms{y}(b)U(a|b)-\frac{1}{n}N(a,b|\bm{x},\bm{y})| \le &\; |P\bms{y}(b)U(a|b)-Q(b)U(a|b)|
+ |Q(b)U(a|b)-\frac{1}{n}N(a,b|\bm{x},\bm{y})|\\ 
\le &\;\delta\bms{y}+\delta\bms{xy}.
\end{align*}
Therefore $\bm{x}\in {\sf T}_{U,\delta^{\prime\prime}}^n(\bm{y})$,
$\delta^{\prime\prime} = \delta\bms{y}+\delta\bms{xy}$,
and thus the number of $1$'s in each column is at most $a$:
\[
a = \exp[n(H(U/Q)+\epsilon^{\prime\prime}].
\]

Theorem~\ref{th:jsl} gives an upper bound on $K$, the number of columns in 
${\cal C}$ and thus the code rate
\[
R = \frac{\log K}{n}.
\]
Since $M=|{\sf T}_{Q,\delta\bms{y}}^n|\le \exp[n(H(Q)+\epsilon\bms{y})]$, we have
\begin{align*}
K & \le \frac{M}{v}(1+\log a)\\
& \le \exp[n(H(Q)-H(W/P)+\epsilon\bms{y}+\epsilon^\prime)]\cdot[1+n(H(U/Q)+\epsilon^{\prime\prime})].
\end{align*}

Now, for any $N\times K$ submatrix of $\bm{A}$ with no all-zero rows,  
we have $K\cdot a \ge N\cdot 1$, and thus
\begin{align*}
K & \ge \frac{N}{a}\\
& \ge \exp[n(H(P)-H(U/Q)+\epsilon\bms{x}-\epsilon^{\prime\prime})].
\end{align*}
Therefore, the compression $R$ is bounded by
\[
I(P,W)+\epsilon\bms{x}-\epsilon^{\prime\prime}\le
R \le I(P,W)+\epsilon\bms{y}+\epsilon^\prime + 
\frac{\log[1+n(H(U/Q)+\epsilon^{\prime\prime})]}{n},
\]
where $\epsilon\bms{x}, \epsilon\bms{y}, \epsilon^\prime, \epsilon^{\prime\prime}\rightarrow 0$,
as $n\rightarrow\infty$, and
the compression rate $I(P,W)$ is asymptotically optimal.

Let us now compare the compression problem in this section with the earlier one in Sec.~\ref{sec:ce}.
In the earlier case, for each Alice's sequence of coins $\bm{x}$, Bob most likely chooses 
one of the approximately $\exp\{nH(W/P)\}$ sequences of faces $W$-generated
by $\bm{x}$, each one with roughly the same probability.
In the case we just considered, for each Alice's $\bm{x}$, Bob's sequence of faces 
will always be a fixed sequence $\bm{y}(\bm{x})$, $W$-generated by $\bm{x}$.
Note that in both cases, after a $\bm{y}\in {\sf T}_{W,\delta}^n(\bm{x})$ has been identified for Bob,
his uncertainty about $\bm{x}$ reduces from $H(P)$ to $H(U/Q)$; hence the same
compression rate.

To an observer who can see only the sequences of faces at both
ends, Alice and Bob now do not appear to be identical sources for any rate of compression
$R$ smaller than the entropy $H(Q)$: Bob has about $\exp(nR)$ different,
equally likely face-sequences
of length $n$ whereas Alice has about $\exp(nH(Q))$.
For each Alice's $\bm{x}$, for quantum transmission or storage, 
Bob can prepare the quantum state $\ops{\Psi\bms{y}}$ instead of 
the sequence of faces $\bm{y}$. In the scenario of Sec.~\ref{sec:ce},
his state is roughly a uniform mixture of pure states
$\ops{\Psi\bm{{}_{y}}}$ where each $\bm{y}$ is $W$-generated by $\bm{x}$,
whereas in the scenario of this section, his state is the pure state $\ops{\Psi\bms{y(x)}}$.

\subsection{Channel Coding and Lossy Mixed State Compression}
The Bhattacharyya distance is in classical information theory most commonly 
known for its role in bounding
the error-probability of a discrete memoryless channel (DMC):
Consider a DMC with input alphabet ${\cal X}$, output alphabet ${\cal Y}$,
and transition probabilities $W(b|a)$, $a\in {\cal X}$, $b\in {\cal Y}$.
When sequence $\bm{x}\in{\cal X}^n$ has been transmitted,
the probability 
that the maximum likelihood detector finds sequence
$\bm{x^{\prime}}\in{\cal X}^n$ more likely is
smaller than 
\[
\sum_{\bm{y}\in {\cal Y}^n}
\sqrt{W^n(\bm{y}|\bm{x})W^n(\bm{y}|\bm{x^{\prime}})}.
\]
This bound is known as the {\it Bhattacharyya bound} and its negative 
logarithm as the Bhattacharyya distance between
sequences $\bm{x}$ and $\bm{x^{\prime}}$.
The probability of error for the maximum likelihood decoder can then be
bounded in terms of the rate of the channel code 
used for transmission. One way to derive such bound is by solving a special
rate distortion problem. We state the problem below and describe its connection
with particular lossy mixed state compression. For its application to channel coding,
we refer the reader to \cite{jo} or textbooks \cite[pp.~185, 193]{CK} and \cite[pp.~408--410]{vo}.

Consider a lossy mixed state compression problem where both the original source Alice and 
the reproduction source Bob have the same alphabet ${\cal X}$.
The fidelity between sequences $\bm{x}$ and $\bm{x^\prime}$ is 
the Bhattacharyya-Wooters overlap between 
$W^n(\cdot|\bm{x})$ and $W^n(\cdot|\bm{x^{\prime}})$:
\begin{equation}
F(\bm{x},\bm{x^{\prime}}) = \Bigl[\sum_{\bm{y}\in {\cal Y}^n}
\sqrt{W^n(\bm{y}|\bm{x})W^n(\bm{y}|\bm{x^{\prime}})}\Bigr]^2.
\label{eq:rdf}
\end{equation}
Let ${\cal C}\subseteq {\cal X}^n$ be a reproduction code. We encode source sequence
$\bm{x}\in{\cal X}^n$ by choosing the codeword $\hat{\bm{x}}$ which maximizes
the fidelity $F(\bm{x},\hat{\bm{x}})$. Let $F(\bm{x}|{\cal C})$ denote this maximum
fidelity:
\[
F(\bm{x}|{\cal C})=\max_{\hat{\bm{x}}\in{\cal C}}F(\bm{x},\hat{\bm{x}}),
\]
and $F({\cal C})$, the expected fidelity achieved with
code ${\cal C}$: 
\[
F({\cal C}) = \sum_{\bm{x}\in{\cal X}^n}P(\bm{x})F(\bm{x}|{\cal C}).
\]
We are interested in finding out how the fidelity $F({\cal C})$ depends on the rate of
code ${\cal C}$.

We get the answer to the question through the following rate distortion problem.
Let again source and reproduction alphabet be ${\cal X}$.
Define a single-letter distortion measure between a source letter $a$
and a reproduction letter $a^{\prime}$ to be the Bhattacharyya
distance between the letters:
\[
d_W(a, a^{\prime}) = -\log\sum_{b\in{\cal Y}}\sqrt{W(b|a)W(b|a^{\prime})},~~
a, a^{\prime}\in {\cal X}.
\]
To make the distortion finite, we shall assume that any two coins 
have at least one common face,
{\it i.e.,} far all $a,a^{\prime}\in{\cal X}$, there is a $b\in{\cal Y}$
such that $W(b|a) > 0$ and  $W(b|a^{\prime}) > 0$. Thus, we have
\[
0\le d_W(a,a^{\prime}) \le d_0,~~
a, a^{\prime}\in {\cal X}.
\]
Because of our assumption that $W$ has no identical rows, 
$d_W(a,a^{\prime})= 0$ iff $a=a^{\prime}$.

The distortion between sequences is the average of the per letter distortion between 
elements of the sequences:
\begin{align*}
d_W(\bm{x},\bm{x^{\prime}}) = 
\frac{1}{n}\sum_{i=1}^n d(x_i, x_i^{\prime}) = &
-\frac{1}{n}\log\prod_{i=1}^n\Bigl[\sum_{b\in{\cal Y}}\sqrt{W(b|x_i)W(b|x_i^{\prime})}\Bigl]\\
 = & -\frac{1}{n}\log\sum_{\bm{y}\in {\cal Y}^n} 
\sqrt{W^n(\bm{y}|\bm{x})W^n(\bm{y}|\bm{x^{\prime}})}, ~~
\bm{x}, \bm{x^{\prime}}\in {\cal X}^n.
\end{align*}
The fidelity (\ref{eq:rdf}) is therefore given by
\[
F(\bm{x},\bm{x^{\prime}}) = 
exp(-2nd_W(\bm{x},\bm{x^{\prime}})).
\]
Note that if the distortion between two sequences remains strictly positive as $n$ increases, 
the fidelity between them approaches 0.

Let ${\cal C}\subseteq {\cal X}^n$ be a reproduction code. We encode source sequence 
$\bm{x}\in{\cal X}^n$ by choosing the codeword $\hat{\bm{x}}$ that minimizes 
the distortion $d(\bm{x},\hat{\bm{x}})$. Let $d(\bm{x}|{\cal C})$ denote this minimum distortion:
\[
d(\bm{x}|{\cal C})=\min_{\hat{\bm{x}}\in{\cal C}}d(\bm{x},\hat{\bm{x}}),
\]
and $d({\cal C})$, the expected distortion achieved with
code ${\cal C}$: 
\[
d({\cal C}) = \sum_{\bm{x}\in{\cal X}^n}P(\bm{x})d(\bm{x},{\cal C}).
\]

Let $V$ be an $|{\cal X}|\times |{\cal X}|$ stochastic matrix with elements
$V_{aa^\prime}=V(a^\prime|a)$, $a,a^\prime\in {\cal X}$, and let
\[
d(V) = \sum_{a,a^\prime\in {\cal X}}P(a)V(a^\prime|a)d(a,a^\prime).
\]
be the {\it average distortion} associated with $V$.
The {\it rate distortion function} of a DMS with generic distribution $P$ is
given by
\[
R(D) = \max_{V:d(V)\le D} I(P,V).
\]
Its significance, found by Shannon in \cite{sh:fc}, is expressed by the 
source coding theorem and its converse (see \cite[pp.~397--400]{vo} for the form used 
here).
Before stating the theorem and its application to our problem, we compute
the distortion measure $d_W(\cdot,\cdot)$ and the rate distortion function for the source
of Example~\ref{ex:bsc}.
\begin{example}
Consider again the source shown in Fig.~\ref{fig:bsc}. We have
\[
d_W(C_1, C_2) = -\log\sqrt{4p(1-p)} ~\text{and} ~
d_W(\bm{x},\bm{x}^\prime) = \frac{1}{n}D_{H}(\bm{x},\bm{x^\prime})\cdot d_W(C_1, C_2),
\]
where $D_{H}(\bm{x},\bm{x}^\prime)$ is the Hamming distance between sequences $\bm{x}$
and $\bm{x^\prime}$. The rate distortion function is given by
\[
R(D)= H(P)-h(D/d_W(C_1,C_2)). 
\]
\end{example}
Note that $R(0) = H(P)$, which is true in general under our assumptions.

\begin{theorem} 
\cite[pp.~397--400]{vo}
{\it Source Coding Theorem and its Converse}\\
For any block length $n$ and rate $R$,
there exists a block code ${\cal C}\subseteq {\cal X}^n$ with average distortion
$d({\cal C})$ satisfying
\[
d({\cal C})\le D + d_0e^{-nE(R,D)},
\]
where $E(R,D)>0$ for $R>R(D)$. Conversely,
no source code for which $d({\cal C}) \le D $ has rate smaller than $R(D)$.
\label{th:rd}
\end{theorem}

We use this result to show how the fidelity $F({\cal C})$ depends on the rate of
code ${\cal C}$:

\begin{theorem}
For any block length $n$ and rate $R>H(P)$, a rate $R$ block code ${\cal C}\subseteq {\cal X}^n$
exists such that the fidelity $F({\cal C})\rightarrow 1$ as $n\rightarrow 0$.
Conversely, for any code ${\cal C}\subseteq {\cal X}^n$ with rate $R<H(P)$,
$F({\cal C})\rightarrow 0$ as $n\rightarrow 0$.
\end{theorem}
\begin{proof}
By the Source Coding Theorem~\ref{th:rd}, we have
\begin{align*}
F({\cal C}) = &\sum_{\bm{x}\in{\cal X}^n}P(\bm{x})F(\bm{x}|{\cal C})\\
= & \sum_{\bm{x}\in{\cal X}^n}P(\bm{x})\exp(2nd_W(\bm{x}|C))\\
\ge & \sum_{\bm{x}\in{\cal X}^n}P(\bm{x})(1-nd_W(\bm{x}|C))\\
\ge & 1 - n d_0 e^{-nE(R,0)}
\end{align*}
where $E(R,0)>0$ for $R>R(0)=H(P)$. Therefore, the fidelity can be made arbitrarily
close to 1 by increasing the block length $n$.

By the Converse to the Source Coding Theorem~\ref{th:rd}, no code for which
$d({\cal C}) \le 0$ has rate smaller then $H(P)$.  Thus for $R<H(P)$, we have
$d({\cal C}) = D > 0$. Therefore, the distortion $d(\bm{x}|{\cal C})$
remains strictly positive as $n$ increases for a probabilistically
large set of sequences $\bm{x}$. Consequently for the same set the fidelity $F\bm{x}|{\cal C})$ approaches 0.
\end{proof}
 


\newpage
\begin{thebibliography}{10}
\bibitem{sc} B.~W.~Schumacher, ``Quantum coding,''
{\it Physical Review A,} vol.~64, 2001.
%
\bibitem{hor:oc} M.~Horodecki, ``Optimal compression for mixed signal states,''
{\it Phys.\ Rev.\ A,} vol.~57, pp.~3364--3369, 1998.
%
\bibitem{hor:cl} M.~Horodecki, ``Limits for compression of quantum information carried
by ensembles of mixed states,''
{\it Phys.\ Rev.\ A,} vol.~61, 052309, 2001.
%
\bibitem{fm} H.~Barnum, C.~M.~Caves, C.~A.~Fuchs, R.~Jozsa, and B.~Schumacher,
``On quantum coding ensembles of mixed states,'' {\it arXiv:quant-ph/0008024.}
%
\bibitem{dvc} W.~D\"{u}r, G.~Vidal, and J.~I.~Cirac, ``Visible compression of commuting mixed
states,'' {\it Phys.\ Rev.\ A,} vol.~51 pp.~2738--2747, 1995.
%
\bibitem{ks} G.~Kramer and S.~.A.~Savari, 
``Quantum data compression of ensembles of mixed states with commuting density operators,'' 
{\it arXiv:quant-ph/0101119.}
%
\bibitem{ki:hms} M.~Koashi and N.~Imoto, ``Compressibility of Mixed-State Signals,''
{\it arXiv:quant-ph/0103128.}
%
\bibitem{sh:mtc}  C.~E.~Shannon, ``A mathematical theory of communication,''
{\it Bell Syst.\ Tech.\ J.,} vol.~27, no.~10, pp.~379--423, 623--656, Oct.~1948.
%
\bibitem{Csiszar98}
I.~Csisz\'{a}r, ``The method of types,'' {\em IEEE Trans.\ Inform.~Theory},
vol.~44, pp.~2505--2523, Oct. 1998.
%
\bibitem{CK}
{I. Csisz\'{a}r and J. K\"{o}rner}, {\em Information Theory: Coding Theorems
  for Discrete Memoryless Systems}.
\newblock Budapest, Hungary: Acad\'{e}miai Kiad\'{o}, 1986.
%
\bibitem{cn} M.~A.~Nielsen and I.~L.~Chuang, {\em Quantum Computation and Quantum Information,}
New York: Cambridge Univ.~Press, 2000.
%
%
%
\bibitem{chll} G.~Cohen, I.~Honkala, S.~Litsyn, and A.~Lobstein, {\it Covering Codes.}
The Netherlands: North Holland, 1997.
%
\bibitem{jo} J.~K.~Omura, ``Expurgated bounds, Bhattacharyya distance, and rate distortion
functions,'' {\it Information and Control,} Vol.~24, pp.~358--383, 1974.
%
\bibitem{vo} A.~J.~Viterbi and J.~K.~Omura, ``Principles of Digital Communication
and Coding,'' New York: McGraw-Hill, 1979.
%
\bibitem{sh:fc} C.~E.~Shannon, ``Coding theorems for a discrete source with a
fidelity criterion,'' {\it IRE National Convention Record,} Part 4, pp.~142--163, 1959.
%
\bibitem{tb} T.~Berger, {\it Rate Distortion Theory: A Mathematical Basis for Data 
Compression,} Englewood Cliffs, NJ: Prentice Hall, 1971.
%
%
%
\end{thebibliography}
\end{document}